\documentclass[aps,prb,twocolumn,superscriptaddress,showpacs,floatfix]{revtex4}

\usepackage{graphicx}
\usepackage{color}
\usepackage{amssymb,amsmath}

\begin{document}

\title{Electronic charge and spin density distribution in a quantum ring \\ 
with spin-orbit and Coulomb interactions}

\author{Csaba Daday}
\affiliation{School of Science and Engineering, Reykjavik University, Menntavegur 1,
IS-101 Reykjavik, Iceland}
\affiliation{Science Institute, University of Iceland, Dunhaga 3, IS-107 Reykjavik,
Iceland}
\author{Andrei Manolescu}
\affiliation{School of Science and Engineering, Reykjavik University, Menntavegur 1,
IS-101 Reykjavik, Iceland}
\author{D. C. Marinescu}
\affiliation{Department of Physics and Astronomy, Clemson University, Clemson, SC 29621,
 USA}
\author{Vidar Gudmundsson}
\affiliation{Science Institute, University of Iceland, Dunhaga 3, IS-107 Reykjavik,
Iceland}

\begin{abstract}
Charge and spin density distributions are studied within a nano-ring structure endowed with Rashba and Dresselhaus spin orbit
coupling (SOI). For a small number of interacting electrons, in the presence of an external magnetic field, the energy spectrum of the system is calculated through an exact numerical diagonalization procedure. The eigenstates thus determined are used to estimate the
charge and spin densities around the ring. We find that when more than two electrons are considered, the
charge-density deformations induced by SOI are dramatically flattened by the Coulomb repulsion, while the spin density ones
are amplified.
\end{abstract}

\pacs{
71.70.Ej, 
73.21.Hb, 
71.45.Lr  
}

\maketitle

\section{Introduction}

The possibility of controlling the flow of the electron spins in
semiconductor structures by external electric means through spin-orbit
interaction (SOI) has dominated the recent past of spintronics
research. This fundamental principle, first explored in the Datta-Das
spin transistor configuration,\cite{datta} has been guiding a sustained
effort in understanding all the phenomenological implications of this
interactions on systems of electrons.  The coupling between spin and
orbital motion results either from the two-dimensional confinement
(Rashba)\cite{rashba} or from the inversion asymmetry of the bulk crystal
structure (Dresselhaus).\cite{dresselhaus} The usual expression of the the spin-orbit
Hamiltonian $H_{SO}$ retains only the linear terms in the electron
momentum ${\bf p}=(p_x,p_y)$ and is given by
\begin{equation}
H_{SO}=\frac{\alpha}{\hbar}(\sigma_x p_y-\sigma_y p_x) +
\frac{\beta}{\hbar}(\sigma_x p_x-\sigma_y p_y) \ .
\label{HSOc}
\end{equation}
The Rashba and Dresselhaus coupling constants are $\alpha$ and
$\beta$, respectively, while $\sigma_{x,y,z}$ are the Pauli matrices.
In general, the two interactions are simultaneously present and often
have comparable strengths.  While $\alpha$,  the coupling constant of the
Rashba interaction, can be modified by external electric fields induced by
external gates, the strength of the Dresselhaus SOI, $\beta$, is fixed by
the crystal structure and by the thickness of the quasi two-dimensional
electron system.\cite{Winkler,Ihn}  In many situations of interest,
an additional energy scale is introduced by the Zeeman interaction of
the electron spin with an external magnetic field, proportional to the
effective gyromagnetic factor, $g^*$, which depends on the material
energy-band structure. While $g^*=-0.44$ is very small in GaAs, it can
be more that 100 times larger in InSb.

The interplay between the two types of SOI, which have competing
effects on the precession of the electron spin as they rotate it
in opposite directions, and the Zeeman splitting, which minimizes
the energy by aligning the spin parallel to the external field,
determines the ground state polarization of the electron system and the
characteristics of spin transport. The investigation of such problems
in mesoscopic rings has been pursued intensively by several authors.
\cite{Splettstoesser,Souma,Sheng,Liu,Nowak}
In particular, it was noticed that, in the absence of the Coulomb
interaction among the electrons, the interference between the Rashba and
Dresselhaus precessions relative to the orbital
motion, leads to the creation of an inhomogeneous spin and charge
distribution around the ring.\cite{Sheng} The charge inhomogeneity
created in this situation has a symmetric structure with two maxima
and two minima around the ring and will be called here a charge-density
deformation (CDD).
The effect of the Coulomb interaction on this type of charge distribution
has been considered for two electrons. It was obtained that, on account
of the electrostatic repulsion, the two electrons become even more
localized in the potential minima associated to the CDD, leading to an
amplitude increase.\cite{Nowak,Liu}

In this work we obtain an estimate of the effect of the Coulomb interaction on the charge and spin distribution associated with $N = 2, 3$ and $4$ electrons in a ring with SOI coupling by an exact diagonalization procedure that uses the configuration interaction method. Our results indicate that when the number of electrons increases, the mutual repulsion leads to more uniform charge distribution around the ring, generating a dramatically flattened
CDD.  In contrast,
the spin-density deformation (SDD) is amplified by the Coulomb effects. This can be explained by the appearance of a stronger repulsion between same spin electrons, leading to more favorable spin orientations.

\section{The Ring model}

\begin{figure}
\includegraphics[width=0.42\textwidth,angle=0,bb=95 56 360 300,clip]{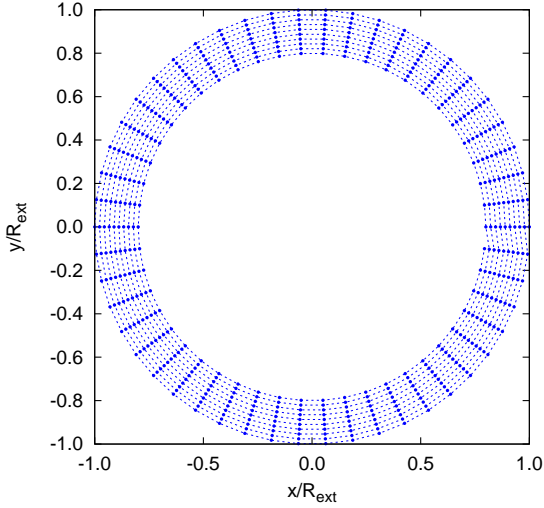}
\caption{(Color online) The discretized ring with $R_{int}=0.8R_{ext}$, and
10 radial $\times$ 50 angular sites. The sites are shown with circular points.
The thin dotted lines connection sites are for guiding the eye.}
\label{Sample}
\end{figure}

The system of interest in our problem is
a two-dimensional quantum ring of exterior and interior radii, $R_{ext}$ and $R_{int}$ respectively. The ring is placed in a perpendicular magnetic field $B$
associated in the symmetric gauge with a vector potential
${\bf A} = B/2(-y,x,0)$.  The single-particle Hamiltonian of an electron of momentum ${\bf
p}=-i\hbar\nabla+e{\bf A}$ and effective mass $m^*$ is written as the sum of an orbital term $H_O={\bf p}^2/2m^*$, a Zeeman contribution $H_Z=(1/2)g^*\mu_B B \sigma_z$ and the spin-orbit coupling given in Eq.~(\ref{HSOc}). The ensuing expression,
\begin{equation}
H=H_O + H_Z + H_{SO} \ ,
\label{Hc}
\end{equation}
is discretized in a standard manner\cite{Splettstoesser,Souma,Meijer} on a grid\cite{footnote} defined by
$N_r$ radial and $N_\varphi$ angular sites, as shown in Fig.\ \ref{Sample}.
The radial coordinate of each site is $r_k=R_{ext}-(k-1)\delta r$, with
$k=1,2,...,N_r$, while $\delta r=(R_{ext}-R_{int})/(N_r-1)$ is the distance
between adjacent sites with the same angle.  Similarly, the angular
coordinate is $\varphi_j=(j-1)\delta\varphi$, where $j=1,2,...,N_{\varphi}$
and $\delta\varphi=2\pi/N_{\varphi}$ is the angle between consecutive sites
with the same radius.
The Hilbert space is spanned by the ket-vectors $| kj\sigma \rangle$,
where the first integer, $k$, stands for the radial coordinate, the
second one, $j$, for the angular coordinate, and $\sigma=\pm 1$ denotes
the spin projection in the $z$ direction.

In this basis $\{ |kj\sigma\rangle \}$, the matrix elements of the orbital
Hamiltonian are given by:
\begin{equation}
\begin{split}
&\langle kj\sigma | H_O | k'j'\sigma'\rangle = \\
& T \delta_{\sigma\sigma'} \Bigg\{ \left[t_\varphi+t_r +
\frac{1}{2}t_B^2\left(\frac{r_k}{4R_{ext}}\right)^2\right]
\delta_{kk'}\delta_{jj'} \\
&-\left[ t_\varphi+t_B\frac{i}{4\delta\varphi} \right] \delta_{kk'}\delta_{jj'+1}
+ t_r \delta_{kk'+1}\delta_{jj'} \Bigg \} + h.c. \ .
\end{split}
\label{HOd}
\end{equation}
$T=\hbar^2/(2m^*R_{ext}^2)$ is the energy unit, while $R_{ext}$ is the length unit.
In  $T$ units, we obtain
$t_\varphi=[R_{ext}/(r_k\delta\varphi)]^2$ the angular hopping energy,
$t_r=(R_{ext}/\delta r)^2$ the radial hopping energy, and $t_B=\hbar
eB/(m^*T)$ the magnetic cyclotron energy. ($h.c.$ denotes the Hermitian
conjugate.)

In the same basis, the Zeeman Hamiltonian is simply diagonal in the
spatial coordinates,
\begin{equation}
\langle kj\sigma | H_Z | k'j'\sigma'\rangle=\frac{1}{2}Tt_B\gamma
(\sigma_z)_{\sigma\sigma'}
\delta_{kk'}\delta_{jj'} \ ,
\end{equation}
where $\gamma=g^*m^*/(2 m_e)$ is the ratio between the Zeeman gap and the cyclotron
energy, $m_e$ being the free electron mass.

For the spin-orbit Hamiltonian we obtain:
\begin{equation}
\begin{split}
& \langle kj\sigma | H_{SO} | k'j'\sigma'\rangle = \frac{1}{2}
Tt_\alpha\Bigg[ t_B \frac{r_k}{4R_{ext}}
(\sigma_r^j)_{\sigma\sigma'} \delta_{kk'}\delta_{jj'} \\
& + i t_\varphi^{1/2} \ \frac{(\sigma_r^j+\sigma_r^{j+1})_{\sigma\sigma'}}{2}
\delta_{kk'}\delta_{jj'+1} -it_r^{1/2} \ (\sigma_\varphi^j)_{\sigma\sigma'}
\delta_{kk'+1}\delta_{jj'} \Bigg] \\
& + Tt_\beta\sum_{k,j} \Bigg[
\sigma_r^j \to (\sigma_\varphi^j)^* \ {\rm and} \ \sigma_\varphi^j \to -(\sigma_r^j)^*
\Bigg] + h.c. \ ,
\end{split}
\label{HSOd}
\end{equation}
where $t_{\alpha}=\alpha/(R_{ext}T)$ and $t_{\beta}=\beta/(R_{ext}T)$
are the two types of spin-orbit relative energies, while
$\sigma_r(\varphi)=\sigma_x\cos\varphi+\sigma_y\sin\varphi$ and
$\sigma_\varphi(\varphi)=-\sigma_x\sin\varphi+\sigma_y\cos\varphi$
are the radial and angular Pauli matrices, respectively.  We used
the slightly shorter notations $\sigma_r^j=\sigma_r(\varphi_j)$ and
$\sigma_\varphi^j=\sigma_\varphi(\varphi_j)$ for the matrices at the
particular angles on our lattice. The Rashba spin-orbit terms are all
included in the first square bracket. The Dresselhaus terms are very
similar to the Rashba ones, being given by the substitutions indicated
in the second square bracket.

The single particle states of the noninteracting Hamiltonian (\ref{Hc}),
$H\psi_a=\epsilon_a\psi_a$, are computed as eigenvalues and eigenvectors
of the matrices (\ref{HOd})-(\ref{HSOd}), $\psi_a(r_k,\varphi_j)=\sum_{\sigma} \Psi_{a,\sigma}(k,j)|\sigma\rangle$.
where $\Psi_{a,\sigma}(k,j)$ are $c-$numbers.

In the basis provided by  $\{\psi_a\}$ the {\it interacting}
many-body Hamiltonian is written in the second quantization as
\begin{equation}
{\cal H} = \sum_a \epsilon_a c^\dagger_a c_a^{}
+ \frac{1}{2}\sum_{abcd} V_{abcd} c^\dagger_a c^\dagger_b c_d^{} c_c^{} \,,
\label{Hmany}
\end{equation}
where $c^\dagger_a$ and $c_a^{}$ are the creation and annihilation operators
on the single-particle state $a$. The matrix elements of the Coulomb potential
$V({\bf r}-{\bf r}')=e^2/(\kappa |{\bf r}-{\bf r}'|)$, $\kappa$ being the
dielectric constant of the material,  are in general give by
\begin{equation}
V_{abcd}=\langle \psi_a(\bf r) \psi_b(\bf r') | V({\bf r}-{\bf r}') |
 \psi_c(\bf r) \psi_d(\bf r') \rangle \ .
\label{Vabcd1}
\end{equation}
In the present discrete model the double scalar product is in fact a
double summation over all the lattice sites and spin labels
\begin{equation}
\begin{split}
V_{abcd} & = Tt_C \sum_{\substack{kj\sigma\\k'j'\sigma'}}
 \Psi_{a,\sigma}^*(kj) \Psi_{b,\sigma'}^*(k'j')
\frac{R_{ext}}{|{\bf r}_{jk}-{\bf r}_{j'k'}|} \\
&  \times  \Psi_{c,\sigma}(kj) \Psi_{d,\sigma'}(k'j')  \ .
\end{split}
\label{Vabcd2}
\end{equation}
The new energy parameter introduced by the Coulomb repulsion
is $t_C=e^2/(\kappa R_{ext}T)$.  In the above summation over the sites,
the contact terms ($k=k',j=j'$) are avoided, as their contribution vanishes
in the continuous limit.

The many-body states $\Phi_{\mu}$ are found by solving the
eigenvalue problem for the Hamiltonian (\ref{Hmany}), $${\cal
H}\Phi_{\mu}=E_{\mu}\Phi_{\mu}\ .$$ A potential solution of the equation
is written in the configuration interaction representation 
\cite{Hawrylak,peeters1,peeters2}
as a linear combination of the non-interacting system
eigenstates (Slater determinants),
\begin{equation}
\Phi_{\mu} = \sum_{\alpha} c_{\mu\alpha}|\alpha\rangle \;,
\label{Phimu}
\end{equation}
with $\{|\alpha \rangle =
|i_1^{\alpha},i_2^{\alpha},...,i_K^{\alpha}\rangle \}$  where
$i_a^{\alpha}=0,1$ is the occupation number of the single-particle state
$\psi_a$ and $K$ is the number of single-particle states considered.
The occupation numbers $i_K^{\alpha}$ are listed in the increasing energy order, so
$\epsilon_K$ is the highest energy of the single-particle state included in
the many-body basis.  For any $|\alpha\rangle$ we have $\sum_a
i_a^{\alpha}=N$, which is the number of electrons in the ring.
It is straightforward to derive the matrix elements of ${\cal
H}_{\alpha\alpha'}$ using the action of the creation and annihilation
operators on the many-body basis.  In practice Eq. (\ref{Phimu}) is
convergent with $K$ for a sufficiently small number of electrons, and
sufficiently small ratio of Coulomb to confinement energy, $t_C$.
This procedure, also known as "exact diagonalization", does not rely
on any mean field description of the Coulomb effects, like Hartree,
Hartree-Fock, or DFT. \cite{Pfannkuche}

To be able to carry the numerical calculations in a reasonable amount of time, we
choose a small ring of radii $R_{ext}=50$ nm and 
$R_{int}=0.8R_{ext}$, containing $N\leq 4$ electrons.  The discretization
grid has 10 radial and 50 angular points (500 sites), as shown in
Fig.\ \ref{Sample}.  Two common semiconductor materials used in the
experimental spintronics are used for the selection of the material constants needed: InAs with $m^*=0.023m_e,\ g^*=-14.9,\
\kappa=14.6$, and estimated (or possible) values for the spin-orbit
interactions $\alpha\approx 20$ and $\beta\approx 3$ meVnm; InSb
with $m^*=0.014m_e, g^*=-51.6, \kappa=17.9$, and $\alpha\approx 50, \
\beta\approx 30$ meVnm.\cite{Winkler,Ihn} The relative energies which
we defined are: for InAs $t_{\alpha}=0.60, \ t_{\beta}=0.09, \ t_C=2.9,
\gamma=-0.17$; for InSb $t_{\alpha}=0.92, \ t_{\beta}=0.55, \ t_C=1.5,
\gamma=-0.36$.  In our calculations we have considered material parameters
somewhere in between these two sets: $t_{\alpha}=0.7, \ t_{\beta}=0.3,
\ t_C=2.2, \gamma=-0.2$

\section{Results}

\subsection{Single particle calculations}

In the absence of the SOI, ($\alpha = \beta = 0 $), the single particle Hamiltonian (\ref{Hc})
shares its eigenstates $\psi_a$ with the $\hat{z}$ components of the angular momentum $L_z$ and spin
$S_z=\hbar\sigma_z/2$.  In the presence of only one type of SOI,
either $\alpha \neq 0$ or $\beta \neq 0$, the Hamiltonian
commutes with the $\hat{z}$ component of the total angular momentum, $L_z+S_z$, which is 
conserved. 
When both $\alpha
\neq 0$ and $\beta \neq 0$, the angular momentum is no longer conserved.
However, $\psi_a$ continue to be eigenstates of the parity operator ${\cal P}
= \Pi \sigma_z$, $\Pi$ being the (three dimensional) spatial inversion
operator.  Indeed, the general Hamiltonian (\ref{Hc}) commutes with ${\cal
P}$, which can be easily verified by using $\Pi{\bf p}=-{\bf p}\Pi$ and the
commutation rules of the Pauli matrices. So in general ${\cal P} \psi_a =
s \psi_a$, and thus the parity $s=\pm 1$ of any state $a$ is conserved,
i.\ e.\ it is independent on the magnetic field.  In particular, when
$\alpha=\beta$ and $g^*=0$, all states become parity-degenerate at any
magnetic field. \cite{Sheng, Nowak, Schliemann} We identify the parity
of the single particle states calculated on our discrete ring
model by looking at the relation $\psi_{a,\sigma}(k,j)= s \sigma
\psi_{a,\sigma}(k,\bar\jmath))$ where $(k,j)$ and $(k,\bar\jmath)$
are diametrically opposed sites, with angular coordinates
$\varphi_{\bar\jmath}=\varphi_j+\pi$.

\begin{figure}
\includegraphics[width=0.42\textwidth,angle=0,bb=81 51 296 302,clip]{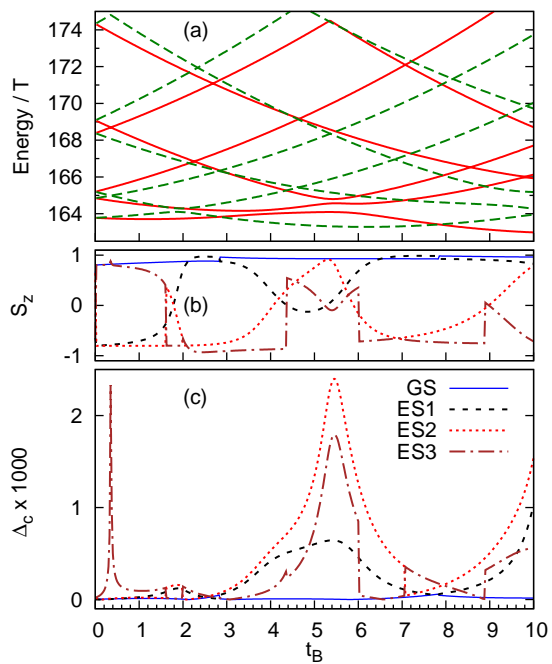}
\caption{(Color online)
(a) The lowest 12 energies of the single particle states vs. the magnetic
energy $t_B$. The solid (red) and the dashed (green) lines show the
states with positive and negative parity, respectively.
(b) The expected value of the spin projection along the $z$ direction
$S_z$, in units of $\hbar/2$, for the first four states on the energy
scale.
(c) The standard deviation $\Delta_c$ of the charge distribution around
the circle with radial site index $k=6$, for the first four energy states:
ground state (GS), first, second, and third exited states (ES1, ES2,
ES3). The same association of line types with states is used in panel
(b). }
\label{N1}
\end{figure}

In Fig.\ \ref{N1} we show the single particle states energy for
$0<t_B<10$ (units of $T$),
which corresponds to a magnetic field strength between 0 and 1.32 Tesla.
Further increment of the magnetic field requires an augmentation of 
the number of sites on the ring in order to maintain the discrete
model as a reasonable approximation of a physically continuous ring.
At zero magnetic field all states are parity degenerate, which is just
the ordinary spin degeneracy.  The degeneracy is in general lifted by a
finite magnetic field. There are, however, some particular values of $t_B$ where the degeneracy persist.  This situation is represented in
Fig.\ \ref{N1}(a) by all intersection points of two lines corresponding to the two possible parities.  Such
intersections do not occur between states with the same parity.
Due to the spin-orbit coupling, the orbital momentum of one state depends
on the spin of the other state and vice versa, and, consequently, states of same parity do in fact interact and thus avoid intersections.\cite{Nowak}
Although in Fig.\ \ref{N1}(a) many states represented by the same line
type apparently cross each other, in reality there are always tiny gaps
between them, similar to those visible at $t_B\approx 2$ between the first and
the second excited states or at $t_B\approx 5.5$ between the first, second,
and third excited states.  The magnitude of these gaps depends
on the $g-$factor, reducing in size for a smaller $g^*$ parameter.

In Fig.\ \ref{N1}(b) the evolution of the expected spin in the
$z$ direction, $S_z= \hbar \langle \psi_a | \sigma_z |  \psi_a \rangle /
2$ is presented for the first four states in the energy order.  One can see how the
spin flips for states avoiding the crossing, like those with negative parity
at $t_B\approx 2$ (dashed and dotted lines in Fig.\ \ref{N1}(b)).

Only one type of SOI, either Rashba or Dresselhaus, is sufficient
to avoid the crossing of states with the same parity, but in this
case the charge and the spin densities are uniform around the ring.
When {\em both} SOI types are present the charge and spin
densities become nonuniform.  This situation is equivalent with the
presence of a potential with two maxima and two
minima around the ring, having reflection symmetry relative to the
axes $y=x$ (or $\varphi=\pi/4$, corresponding to the crystal direction
$[110]$) and $y=-x$ (or $\varphi=-\pi/4$, corresponding to the crystal
direction $[1\bar 1 0]$).\cite{Sheng,Nowak}  The amplitude of
the CDD is illustrated in Fig.\ \ref{N1}(c) where the standard deviation 
(in the statistical sense) $\Delta_c$
of the charge density calculated around one circle on the ring, close to
the mean radius, with radial site index $k=6$, is plotted for the lowest four
energy states.  The density deformation occurs on account of the two combined
SOI types which lead to spin interference and additional interaction
between states with the same parity.  Consequently, the amplitude of the
CDD for a certain state is maximum at those magnetic fields where the parity 
degeneracy is lifted (the state avoids a crossing with another state of the same parity).
In Fig.\ \ref{N1}(c) this is clearly seen at $t_B\approx 2$, for the
excited state.  In this example the CDD in the ground state is very weak.
The sharp peak at $t_B\approx 0.4$ indicates the existence of a narrow gap between the 4-th and 5-th energy states that avoid crossing.

\subsection{Many particle calculations}

\begin{figure}
\includegraphics[width=0.48\textwidth,angle=0,bb=50 50 300 304,clip]{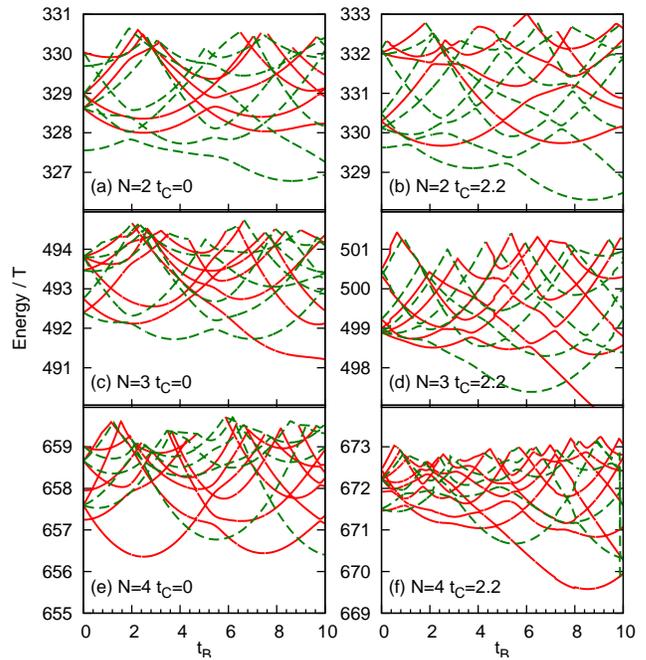}
\caption{(Color online) Energy spectra of the first 12 states for
$N=2,3$, and 4 electrons without Coulomb interaction, $t_C=0$ in
panels (a),(c),(e), and with Coulomb interaction, $t_C=2.2$, in panels
(b),(d),(f).  The solid (red) and the dashed (green) lines show the
states with positive and negative parity, respectively.
}
\label{EN}
\end{figure}

In the following considerations, we will include more than one electron. In Fig.\ \ref{EN}
we compare the energy spectra for the first 12 states vs. the magnetic energy
for $N=2,3$, and $4$ electrons, without and with Coulomb interaction.
Since the Coulomb interaction is invariant at spatial inversion (and
independent on spin) the parity $s$ is also conserved in the many-body
states. Spectra drawn for $t_C=0$ and $t_C=2.2$ have similar features.  The interacting spectrum presents a shift to higher energies,
on account of the additional Coulomb energy, and a slight increase of the gaps
at high magnetic fields.  Moreover, the
crossings and the anti-crossings (points where the crossings were avoided) of the energy levels have a tendency to
shift slightly to higher magnetic fields.  Similarly, the gap between the ground
state and the excited states increases at high $t_B$.
\begin{figure}
\includegraphics[width=0.48\textwidth,angle=0,bb=50 50 300 300,clip]{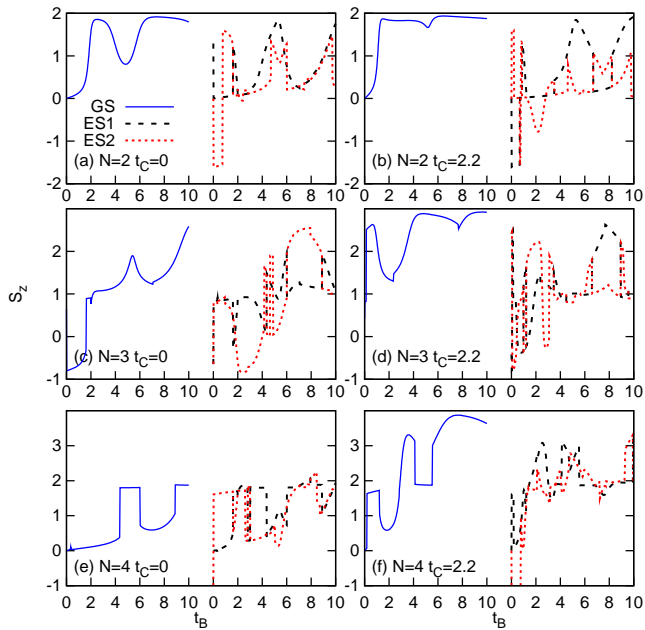}
\caption{(Color online) The total spin projection in the $z$ direction,
in units of $\hbar/2$, for the many body states with $N=2,3,4$ electrons.
Without interaction, i.\ e.\ $t_C=0$ in panels (a),(c),(e), and with
interaction, with $t_C=2.2$, in panels (b),(d),(f).  Only the first
three states are shown here, the ground state (GS), the 1-st excited states
(ES1), and the 2-nd excited state (ES2).  The magnetic energy $t_B$
varies between 0 and 10 and the lines showing the excited states are
shifted to the right, for clarity.
}
\label{SN}
\end{figure}

The total spin $S_z$ for each of the first three energy states is shown in Fig.\
\ref{SN}.  At zero magnetic field, for an even number of
electrons, here $N=2$ or $N=4$, the ground state is non-degenerate and
has total spin $S_z=0$, i.\ e.\ the spin-up and spin-down states of
individual electrons compensate. When the field is applied, the first spin flip in the
interacting ground state, as well as the spin saturation, occur at lower magnetic fields than in the absence of the Coulomb repulsion. This is a result of the mixing of spin states with the
same parity produced by the interaction.
For $N=3$ the ground state is spin
(double) degenerate at zero field.  In the presence of the Coulomb
interaction, the total spin in the ground state and the higher state is reversed.  

\begin{figure}
\includegraphics[width=0.48\textwidth,angle=0,bb=50 50 300 300,clip]{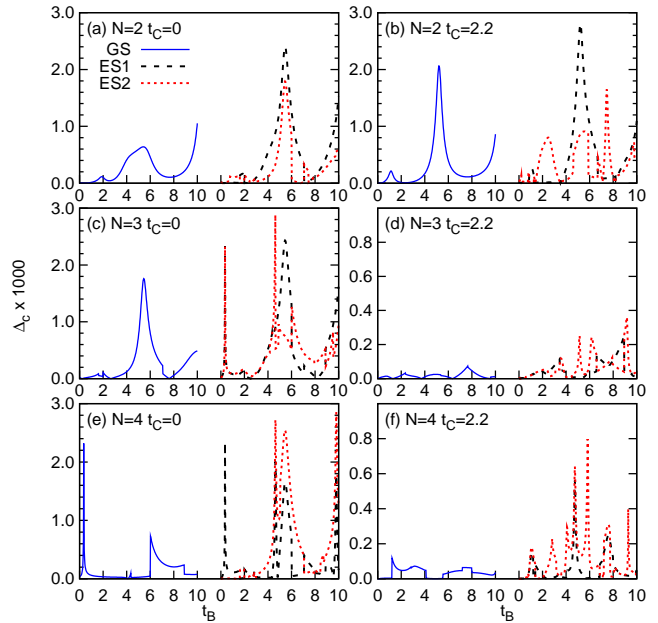}
\caption{(Color online) The standard deviation $\Delta_c$ of the charge
on the circle $k=6$ around the ring, as a measure of the amplitude of the
charge deformation. Shown are the results for the ground state (GS), the
1-st excited states (ES1), and the 2-nd excited state (ES2), for $N=2,3,4$
electrons without without ($t_C=0$), and with interaction ($t_C=2.2$).
The amplitude of the CDD's is strongly reduced by the Coulomb effects
for $N=3,4$; notice the different scales used in the paired panels (c),(d)
and (e),(f).  The magnetic energy $t_B$ varies between 0 and 10 and the
lines corresponding to the excited states are shifted to the right,
for clarity.
}
\label{CDWN}
\end{figure}
\begin{figure}
\includegraphics[width=0.48\textwidth,angle=0,bb=50 50 300 300,clip]{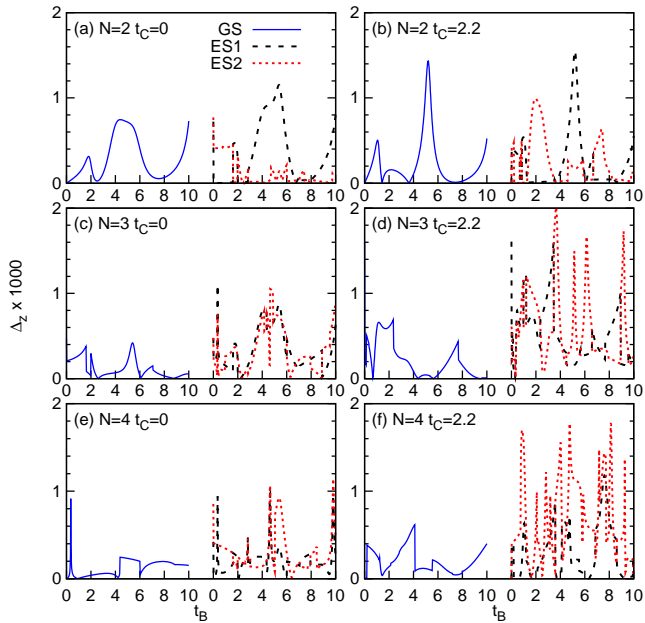}
\caption{(Color online) The standard deviation $\Delta_z$ of the spin
projection along the $z$ direction on the circle $k=6$ around the ring, as
a measure of the amplitude of the spin-density wave.  Like in the previous
figures GS, ES1, and ES2 in the legend indicate the ground state, the 1-st
excited states, and the 2-nd excited states, respectively.  The results
are shown for $N=2,3,4$ electrons without ($t_C=0$), and with
interaction ($t_C=2.2$).  Unlike the CDD's, the SDD's are amplified by
the Coulomb interactions for all $N$.  The magnetic energy $t_B$ varies between
0 and 10 and the lines corresponding to the excited states
are shifted to the right, for clarity.
}
\label{SDWN}
\end{figure}
\begin{figure}
\includegraphics[width=0.48\textwidth,angle=0,bb=51 134 278 272,clip]{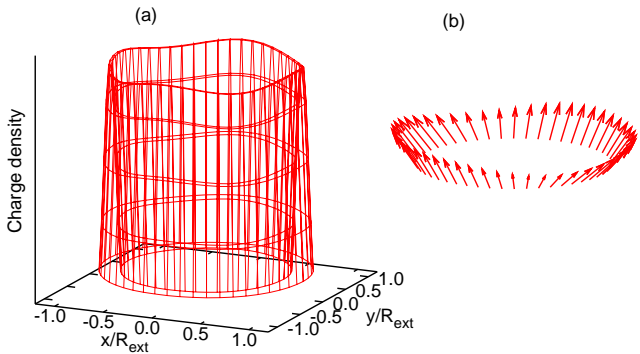}
\caption{(Color online) (a) The charge density for $N=4$ electrons
with interaction ($t_C=2.2$), in the second excited state, i.\ e. ES2 in
Fig.\ref{CDWN}(f), with magnetic energy energy $t_B=4.5.$ (b) The corresponding
total spin distribution along the ring $k=6$ where the standard deviation of the
$z$ component is calculated and shown in Fig.\ref{SDWN}(f).
}
\label{CSd}
\end{figure}

Similar to the case of one electron ($N=1$, Fig.\ \ref{N1}), the charge
deformation of each state is maximized for those magnetic fields where
the state has an anti-crossing (or repulsion) with another state of
the same parity.  The charge deformation parameter $\Delta_c$ is shown
in Fig.\ \ref{CDWN}. For $N=2$ the amplitude of the CDD increases with the Coulomb interaction.  There is a simple reason for
that: the potential associated with the charge deformation has two minima
diametrically opposite on the ring and each of the two electrons tends
be localized in one of these minima. The mutual Coulomb repulsion fixes the electrons in those places better.\cite{Nowak,Liu}  The
situation changes, however, for $N>2$.  The Coulomb forces spread the electrons
differently, more or less uniformly, such that the charge deformation
created by the SOI is drastically reduced.  In other words, the associated
potential is strongly screened even by one extra electron above $N=2$.
This effect can be clearly seen in Fig.\ \ref{CDWN}, comparing panels
(c) with (d) and (e) with (f). The vertical scale of panels (d) and (e)
has been magnified three times, for visibility.

In principle, the screening of the charge deformation is not particularly
related to the spin-orbit effects.  SOI only generates the specific
effective potential which determines the CDD.  In the absence
of SOI ($\alpha=\beta=0$), we checked that a similar screening effect
occurs in the presence of a potential that induces a charge deformation
comparable to that obtained with the SOI.  It is, however, surprising that by adding only one
extra electron such a drastic effect ensues.

Next, we investigate the effect of the Coulomb interaction on the spin
distribution around the ring. The standard deviation of
the spin density projected along the $z$ direction, $\Delta_z$, is plotted in 
Fig.\ \ref{SDWN} where we show the results calculated as before for
the circle corresponding to the sites with radial coordinate $k=6$.
The spin density deformation (SDD) is actually amplified by the Coulomb
interaction for all $N=2,3,4$.  As the CDD's, the SDD's reach their maximum at
the magnetic fields where level repulsion occurs and remains prominent
even when the gaps are very small. The Coulomb enhancement is a result
of the mixing of states with the same parity, but with different spin
orientation produced by the Coulomb potential. Consequently the SDD's
have in general a richer structure than the CDD's.

Finally, in Fig.\ \ref{CSd} we display an example of CDD and SDD,
obtained for $N=4$ interacting particles.
The charge and spin distributions corresponding to the second
excited state and $t_B=4.5$ are illustrated in Figs.\ \ref{CDWN}(f) and \ref{SDWN}(f),
respectively.  The CDD is weak, but still it has four visible maxima.  For two
electrons the CDD has only two maxima which are along the directions $x=y$
or $x=-y$, depending on the state and on the magnetic field, both with
and without the Coulomb interaction.  In particular, for a strictly
one-dimensional ring model and $N=2$, in the ground state, the maxima are
always along the line $x=-y$, \cite{Sheng,Liu} whereas for a two-dimensional
model they can also be along $x=y$.\cite{Nowak}  But in general, for
$N>2$ electrons, screening effects may distribute the charge in more
complicated configurations.  Similar profiles with multiple local
oscillation may be obtained for the spin density, eventually becoming
spin-density waves around the ring.

\section{Conclusions}

We calculated the many-body states of a system of $N=2,3$, and $4$
interacting electrons located in a ring of finite width with Rashba and
Dresselhaus spin-orbit coupling, in the presence of a magnetic field
perpendicular on the surface of the ring.  The Coulomb effects are fully
included in the calculation via the "exact diagonalization" method.
We obtained inhomogeneous charge densities, or CDD's, around the ring
due to the combined effect of the two types of SOI.  When the Coulomb
interaction is included the charge deformation is amplified for $N=2$,
as also shown by other authors.\cite{Nowak,Liu}  For $N>2$ we find
that the CDD is dramatically flattened out in the presence of the
Coulomb interaction.  We interpret the result as a screening effect.
On the contrary, the spin inhomogeneities, or SDD's, are amplified by
Coulomb effects for all $N>1$.

\begin{acknowledgments}
This work was supported by the Icelandic Research Fund.  Valuable discussions
with Sigurdur Erlingsson, Gunnar Thorgilsson, and Marian Ni\c t\u a  are cordially
acknowledged.
\end{acknowledgments}



\begin{thebibliography}{15}
\bibitem{footnote}
The first and second derivatives of any generic function $f(x)$,
are approximated as $f'(x)\approx \left[ f(x+h)-f(x-h) \right]/h$
and $f''(x)\approx \left[ f(x+h)+f(x-h)-2f(x) \right]/h^2$, respectively,
where $h$ is considered sufficiently small.
\end{thebibliography}

%

\end{document}